\documentclass[journal,transmag]{IEEEtran}

\usepackage{cite}
\usepackage{graphicx}
\usepackage{amssymb,amsmath}
\usepackage{hyperref}
\usepackage{widetext}
\usepackage{color}

\begin{document}

\bstctlcite{IEEEexample:BSTcontrol}

\title{Electrically controllable exchange bias via interface magnetoelectric effect}


\author{
	\IEEEauthorblockN{Adam B. Cahaya\IEEEauthorrefmark{1,2,*}, Ansell Alvarez Anderson\IEEEauthorrefmark{1}, Anugrah Azhar\IEEEauthorrefmark{3,4} and Muhammad Aziz Majidi\IEEEauthorrefmark{1}}
	\IEEEauthorblockA{\IEEEauthorrefmark{1}Department of Physics, Faculty of Mathematics and Natural Sciences, Universitas Indonesia, Depok 16424, Indonesia}
	\IEEEauthorblockA{\IEEEauthorrefmark{2}Research Center for Quantum Physics, National Research and Innovation Agency (BRIN), South Tangerang 15314, Indonesia}
	\IEEEauthorblockA{\IEEEauthorrefmark{3}	Department of Physics and Astronomy, The University of Manchester, Oxford Road, Manchester M13 9PL, United Kingdom.}
	\IEEEauthorblockA{\IEEEauthorrefmark{4}Physics Study Program, Syarif Hidayatullah State Islamic University Jakarta, South Tangerang 15412, Indonesia}
	\IEEEauthorblockA{\IEEEauthorrefmark{*}adam@sci.ui.ac.id}	
}

\IEEEtitleabstractindextext{%
\begin{abstract}
Exchange bias is a unidirectional magnetic anisotropy that often arise from interfacial interaction of a ferromagnetic and antiferromagnetic layers. In this article, we show that a metallic layer with spin-orbit coupling can induces an exchange bias via an interface magnetoelectric effect. In linear response regime, the interface magnetoelectric effect is induced by spin-orbit couplings that arises from the broken symmetry of the system. Furthermore, we demonstrate that the exchange bias can be controlled by electric field.
\end{abstract}

}

\maketitle

\pagestyle{empty}
\thispagestyle{empty}

\IEEEpeerreviewmaketitle

\section{Introduction}

Magnetization manipulation in magnetic memory is one of the objectives of spintronics research area \cite{Dieny2020}. Due to low power consumption of voltage-driven magnetization
dynamics \cite{Nozaki_2012}, electrical control of magnetization has the potential for a more efficient manipulation of magnetic memories \cite{CHEN2022169753,SONG201733}. The mechanism to couple voltage and magnetization includes the control of exchange bias using electric field \cite{PhysRevLett.110.067202}.

Exchange bias has attracted much attention due to its applications on magnetic sensors and spintronic devices \cite{C7NR05491B}. In magnetic heterostructure, exchange bias is a unidirectional anisotropy that can occur due to the hard magnetization behavior of an adjacent antiferromagnet \cite{NOGUES1999203}. The anisotropic exchange interaction at the interface of antiferromagnet and ferromagnet induces an exchange bias on the magnetization of the system \cite{Stamps_2000}. The signature of exchange bias is the shift of the center of magnetic hysteresis loop from the origin\cite{KIWI2001584}. The manipulation of exchange bias in magnetic heterostructure motivates innovative designs for spintronics devices \cite{Lin_2019}. The exchange bias in magnetic heterostructure is mediated by the conduction electron of a neighboring metallic layer \cite{PhysRevLett.79.4270}. The spin-orbit coupling due to a non-magnetic metallic layer can be utilized for manipulating exchange bias of the magnetic heterostructure \cite{Peng_2020}. 

The spin-orbit coupling may arise from noncentrosymmetry of the bulk \cite{PhysRevB.96.115204} and Rashba effect at the interface \cite{Lin2019,Bihlmayer_2015}. Furthermore, materials with noncentrosymmetry structure \cite{Sakhnenko_2012} and Rashba effect \cite{PhysRevLett.109.226804} has been shown to have magnetoelectric effect. The spin–orbit coupling leads to spin-dependent electric dipole moments of the electron orbitals, which results in magnetoelectric effect \cite{Sakhnenko_2012}. Magnetoelectric effect enables electric control of magnetic phase of multiferroic materials \cite{Fiebig_2005}. Moreover, magnetoelectric effect in a multiferroic heterostructure can lead to ultralow power magnetic memory \cite{Fujii2022}. 

Here we show that the spin-orbit coupling at the interface can induces an exchange bias in the neighboring ferromagnetic layer via an interface magnetoelectric effect. Sec.~\ref{SecSOI} discusses the linear response theory of interface magnetoelectric effect in a system with linear spin-orbit coupling due to noncentrosymmetry structure and Rashba effect. Sec.~\ref{SecExchangeBias} discusses the exchange bias that arise from the interface magnetoelectric effect. Lastly, Sec.~\ref{SecSummary} summarizes our findings.

\section{Interface magnetoelectric effect due to spin-orbit coupling}
\label{SecSOI}

In second quantization, the interactions in the metallic system near the interface can be written with the following Hamiltonian 
\begin{align}
H=&H_0+H_{\rm int},\label{Eq.Hamiltonian}\notag\\
H_0=&\sum_{\textbf{k}\beta\gamma} a_{\textbf{k}\beta}^\dagger a_{\textbf{k}\gamma} \left[\varepsilon_k\delta_{\beta\gamma}+\alpha_b\boldsymbol{\sigma}_{\beta\gamma}\cdot\textbf{k}+\alpha_R\boldsymbol{\sigma}_{\beta\gamma}\cdot\left(\textbf{k}\times\hat{\textbf{z}}\right)\right]\notag,\\
H_{\rm int}=&\int d^3r\left[\rho(\textbf{r})\phi (\textbf{r})-\textbf{M}(\textbf{r})\cdot \textbf{B}(\textbf{r}) \right].
\end{align}
Here $H_0$ is the unperturbed Hamiltonian \cite{PhysRevB.96.115204}. $H_{\rm int}$ is the interaction Hamiltonian, which represents the potential energy of electric charge density $\rho$ due to electric potential $\phi$ and magnetization $\textbf{M}$ due to magnetic field $\textbf{B}$ \cite{Griffiths1998-ov}.
$a_{j\beta}^\dagger (a_{j\beta})$ is the creation (annihilation) operator of conduction electron with wave vector $\textbf{k}$ and spin $\beta$,  $\boldsymbol{\sigma}$ is Pauli vectors, $\varepsilon_\textbf{k}=\hbar^2k^2/2m$ is the energy dispersion of conduction electron. $\hat{\textbf{z}}$ is normal to the interface. $\alpha_b$ is the coupling constant of spin-orbit coupling of noncentrosymmetric metals \cite{PhysRevB.96.115204,PhysRevX.5.011029}. $\alpha_R$ is the coupling constant of Rashba spin-orbit coupling due to broken symmetry at the interface \cite{Rashba1,Rashba2}. The spin-orbit coupling strength usually $\alpha_{b,R}k_F\sim 0.01$ eV.

Magnetoelectric effect focuses on how magnetization 
\begin{align}
\textbf{M}(\textbf{r})=&-\mu_B\sum_{\textbf{kq}\beta\gamma}e^{i\textbf{q}\cdot\textbf{r}} a_{\textbf{k}+\textbf{q}\beta}^\dagger \boldsymbol{\sigma}_{\beta\gamma} a_{\textbf{k}\gamma}\equiv-\mu_B\textbf{s}(\textbf{r}) \label{Eq.m_s}
\end{align}
 and electric polarization densities $\textbf{P}$ are coupled to magnetic $\textbf{B}$ and electric $\textbf{E}=-\nabla\phi(\textbf{r})$. $\mu_B$ is the Bohr magneton. Here, $\textbf{P}$ is related to charge density
\begin{align}
\rho(\textbf{r})=-\nabla\cdot\textbf{P}=-e\sum_{\textbf{kq}\beta} e^{i\textbf{q}\cdot\textbf{r}} a_{\textbf{k}+\textbf{q}\beta}^\dagger a_{\textbf{k}\beta}\equiv -es_0(\textbf{r}). \label{Eq.n_p}
\end{align}
$\textbf{M}$ and $\textbf{P}$ due to $\textbf{B}$ and $\textbf{E}$ can be determined using linear response theory, in term of charge-spin response matrix $X$
\begin{align}
\left[
\begin{array}{c}
s_0(\textbf{r},t)\\
s_x(\textbf{r},t)\\
s_y(\textbf{r},t)\\
s_z(\textbf{r},t)
\end{array}
\right]=&\int d^3r'dt' X(\textbf{r}-\textbf{r}',t-t')\left[
\begin{array}{c}
-e\phi(\textbf{r},t')\\
-\mu_B B_x(\textbf{r},t')\\
-\mu_B B_y(\textbf{r},t')\\
-\mu_B B_z(\textbf{r},t')
\end{array}
\right]\notag\\
\left[
\begin{array}{c}
s_0(\textbf{q},\omega)\\
s_x(\textbf{q},\omega)\\
s_y(\textbf{q},\omega)\\
s_z(\textbf{q},\omega)
\end{array}
\right]=&X(\textbf{q},\omega)\left[
\begin{array}{c}
-e\phi(\textbf{q},\omega)\\
-\mu_B B_x(\textbf{q},\omega)\\
-\mu_B B_y(\textbf{q},\omega)\\
-\mu_B B_z(\textbf{q},\omega)
\end{array}
\right],\label{Eq.sXB}
\end{align}
where $f(\textbf{q},\omega)$ is the Fourier transform of $f(\textbf{r},t)$. The $jk-$component of $X$ is
\begin{align}
X_{jk}(\textbf{r}-\textbf{r}',t-t')=\frac{i}{\hbar}\theta(t-t')\left[s_j(\textbf{r},t),s_k(\textbf{r}',t')\right].
\end{align} 
One can see that $X_{00}$ is related to the electric susceptibility. $X_{jk}$ ($j,k\neq0$) is the magnetic susceptibility \cite{Cahaya2020ParamagneticAD,PhysRevB.105.214438,CAHAYA2022168874}, its diagonal terms induces an anisotropic response \cite{Cahaya_2021,PhysRevB.103.094420}. $X_{0j}$ and $X_{j0}$ ($j\neq0$) is related to the magnetoelectric susceptibility.

By evaluating the time derivative of $X$ using the unperturbed terms in \ref{Eq.Hamiltonian}, one can show that the Fourier transform of $X(\textbf{r},t)$ is
\begin{align}
X_{jk}(\textbf{r},t)=\delta_{jk}\sum_{\textbf{k}}\frac{f_{\textbf{k}}-f_{\textbf{k}+\textbf{q}}}{\varepsilon_{\textbf{k}+\textbf{q}}-\varepsilon_\textbf{k}+\hbar\omega+i\tau^{-1}}+\delta X_{jk},
\end{align}
where $f_\textbf{k}$ is the Fermi - Dirac distribution for electron with energy $\varepsilon_\textbf{k}$, $\tau\to\infty$ is scattering time. $\delta X_{jk}$ is the linear order correction due to the spin - orbit coupling

\begin{figure}[h]
\centering
\includegraphics[width=0.8\columnwidth]{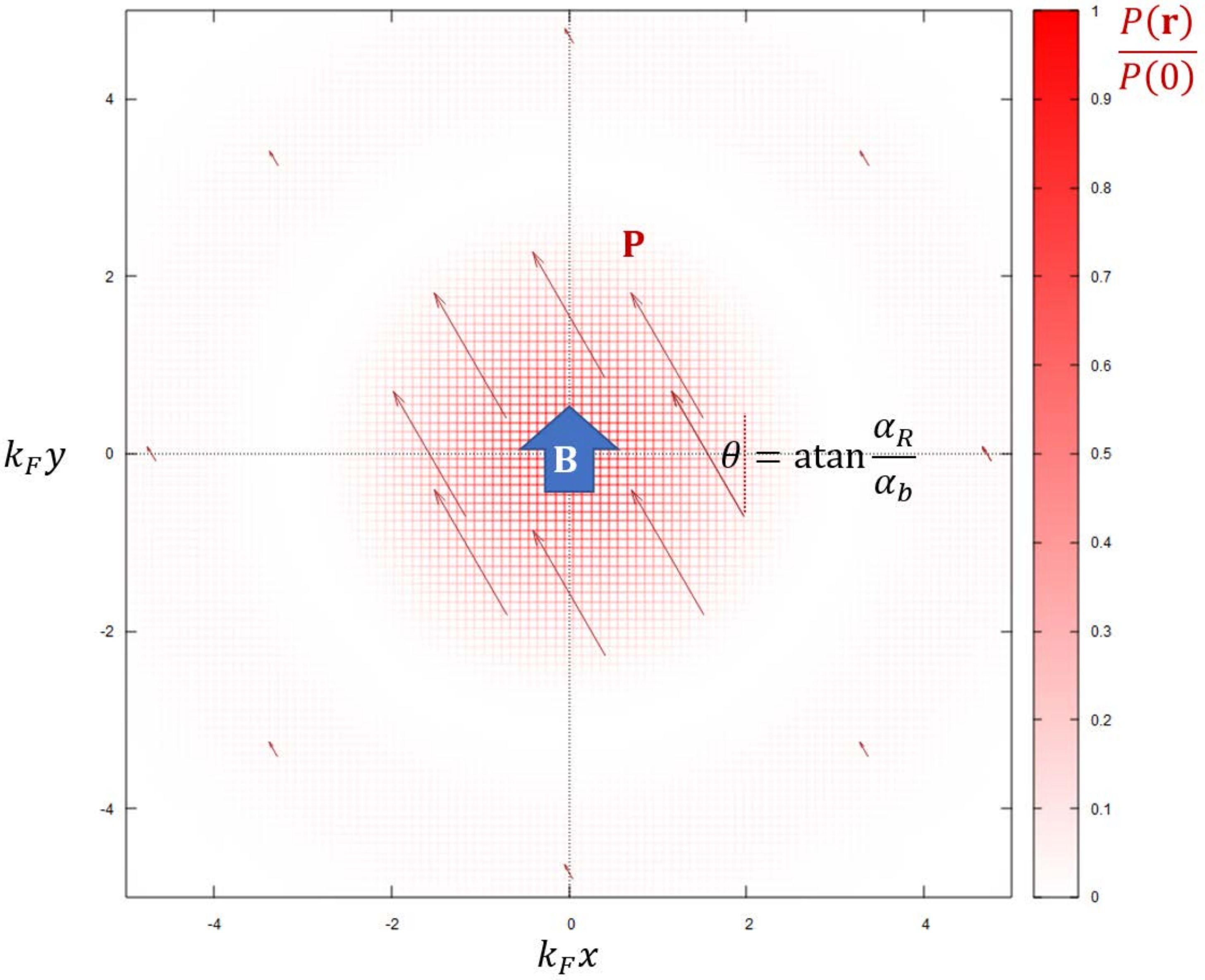}
\caption{A localized magnetic field $\textbf{B}$ at the origin induces electric polarization $\textbf{P}$. The relative angle is determined by the ratio of spin-orbit coupling strengths $\alpha_R$ and $\alpha_b$. The direction and magnitude are indicated by arrow and color. The polarization is localized due to the localization of $x^{-2}\sin^2 x$ function.}
\label{Fig.dipole}
\end{figure}

\begin{widetext}
\begin{align}
&\delta X(\textbf{q},\omega)=\sum_{\textbf{k}}\frac{f_{\textbf{k}}-f_{\textbf{k}+\textbf{q}}}{(\varepsilon_{\textbf{k}+\textbf{q}}-\varepsilon_\textbf{k}+\hbar\omega+i\tau^{-1})^2}\notag\\
&\cdot\left[
\begin{array}{cccc}
0 
& -\alpha_bq_x -\alpha_Rq_y
& -\alpha_bq_y +\alpha_Rq_x
& -\alpha_bq_z \\
-\alpha_bq_x -\alpha_Rq_y
& 0 
& -i\alpha_b(2k_z+q_z)
& i\alpha_b(2k_y+q_y)-i\alpha_R(2k_x+q_x) \\
-\alpha_bq_y +\alpha_Rq_x
& i\alpha_b(2k_z+q_z)
& 0 
& -i\alpha_b(2k_x+q_x)-i\alpha_R(2k_y+q_y) \\
-\alpha_bq_z 
& -i\alpha_b(2k_y+q_y)+i\alpha_R(2k_x+q_x) 
& i\alpha_b(2k_y+q_y)+i\alpha_R(2k_x+q_x) 
& 0 
\end{array}
\right]. \label{Eq.dX}
\end{align}
Substituting \ref{Eq.dX} to \ref{Eq.sXB}, one can show that $\alpha_b$ and $\alpha_R$ generate symmetric and antisymmetric responses, respectively
\begin{align}
\left[
\begin{array}{c}
\textbf{P}(\textbf{q})\\
\textbf{M}(\textbf{q})
\end{array}
\right]
=
\sum_{\textbf{k}}\frac{i\left(f_{\textbf{k}}-f_{\textbf{k}+\textbf{q}}\right)}{(\varepsilon_{\textbf{k}+\textbf{q}}-\varepsilon_\textbf{k}+i\tau^{-1})}\left[
\begin{array}{cc}
\frac{3e^2}{q^2}+\mathcal{O}(\alpha_b,\alpha_R)& \displaystyle\frac{e\mu_B \left(-\alpha_b-\alpha_R \hat{\textbf{z}}\times\right)}{(\varepsilon_{\textbf{k}+\textbf{q}}-\varepsilon_\textbf{k}+i\tau^{-1})}\\
\displaystyle\frac{e\mu_B \left(-\alpha_b+\alpha_R \hat{\textbf{z}}\times\right)}{(\varepsilon_{\textbf{k}+\textbf{q}}-\varepsilon_\textbf{k}+i\tau^{-1})}&\mu_B^2+\mathcal{O}(\alpha_b,\alpha_R)
\end{array}
\right]
\left[
\begin{array}{c}
\textbf{E}(\textbf{q})\\
\textbf{B}(\textbf{q})
\end{array}
\right]
\end{align}
\end{widetext}

When there is no spin-orbit coupling, $\textbf{B}$ and $\textbf{E}$ only responsible for $\textbf{M}$ and $\textbf{P}$, respectively. 
When $\alpha_b,\alpha_R\neq 0$, one find that $\textbf{E}$ and $\textbf{B}$ also generates $\textbf{M}$ and $\textbf{P}$, respectively
\begin{align}
\textbf{P}(\textbf{r})
&=\int d^3r' \frac{e\mu_Bm^2}{2\hbar^4\pi^3} \frac{\sin^2k_F\left|\textbf{r}-\textbf{r}'\right|}{\left|\textbf{r}-\textbf{r}'\right|^2} \left(-\alpha_b-\alpha_R \hat{\textbf{z}}\times\right) \textbf{B}(\textbf{r}'),\notag\\
\textbf{M}(\textbf{r})
&=\int d^3r' \frac{e\mu_Bm^2}{2\hbar^4\pi^3} \frac{\sin^2k_F\left|\textbf{r}-\textbf{r}'\right|}{\left|\textbf{r}-\textbf{r}'\right|^2} \left(-\alpha_b+\alpha_R \hat{\textbf{z}}\times\right)\textbf{E}(\textbf{r}') 
. \label{Eq.diagonalME}
\end{align}
We can see that the polarizations due to bulk spin-orbit coupling contribution are parallel to the field. On the other hand, the polarizations induced by interface Rashba spin-orbit coupling are perpendicular to the fields. The relative angle of polarizations to the fields is determined by the ratio of interface and bulk spin-orbit coupling strength. Fig.~\ref{Fig.dipole} illustrates electric polarization densities due to a localized in-plane magnetic fields $\textbf{B}=B\hat{\textbf{y}}$, which can arise from an exchange interaction with a localized spin. The induced polarization is localized due to the localization of $x^{-2}\sin^2 x$ function. Because of that, it can be assumed that the leading terms of the polarizations is weakly influenced by the periodicity of the system. To avoid divergences when integrated over large volume, $x^{-2}\sin^2 x$ will be approximated using its steepest descent \cite{IPI2539221} 
$$\frac{\sin^2 k_Fr}{(k_Fr)^2}\approx e^{-k_F^2r^2/3}=\int \frac{d^3q}{(2\pi)^3}e^{i\textbf{q}\cdot\textbf{r}}\frac{(3\pi)^{3/2}e^{-3q^2/4k_F^2}}{k_F^3} .$$

\section{Exchange bias due to interface magnetoelectric effect}
\label{SecExchangeBias}

In this section, we focus on a bilayer system that consists of a magnetic layer and a non-magnetic metallic layer. Near the interface, there is a localized magnetic field \cite{PhysRevB.96.144434}
\begin{equation}
\textbf{B}(\textbf{r})=J\textbf{M}\sum_{n}\delta^3(\textbf{r}-\textbf{r}_n)
\label{Eq.Blocal}
\end{equation}
due to the $s-d$ exchange interaction between localized magnetic moment $\textbf{M}$ at position $\textbf{r}_n$ and the spin of conduction electron. $J=2\mu_0/3$ is the exchange constant, which can be estimated from the localized term of dipolar magnetic field \cite{Kutzelnigg1988,CahayaHyper}.
\begin{equation*}
B_{\rm dipolar}= -\frac{2\mu_0}{3}\textbf{M}\delta(\textbf{r})-\mu_0\frac{3\left(\textbf{M}\cdot \hat{\textbf{r}}\right)\hat{\textbf{r}}-\textbf{M}}{4\pi r^3},
\end{equation*}
$\mu_0$ is the vacuum permeability.

Substituting \ref{Eq.Blocal} to \ref{Eq.diagonalME}, we can find that $\textbf{P}$ depends only on $z$ 
\begin{align}
\textbf{P}(\textbf{r})&\simeq -\left[\alpha_b+\alpha_R \hat{\textbf{z}}\times\right]J\textbf{M}\sum_n \frac{e\mu_Bm^2k_F^2}{2\hbar^4\pi^3} e^{-k_F^2\left|\textbf{r}-\textbf{r}_n\right|^2/3} \notag\\
&=-\left[\alpha_b+\alpha_R \hat{\textbf{z}}\times\right]J\textbf{M} \frac{3e\mu_Bm^2N}{2\hbar^4\pi^2 A}e^{-k_F^2z^2/3}, \label{Eq.diagonalME1}
\end{align}
where $N/A$ is number of magnetic moment per unit area. Additionally, from \ref{Eq.diagonalME} we can see that a uniform electric field $\textbf{E}$ induces a uniform $\textbf{M}$
\begin{align}
\textbf{M}(\textbf{r})&\simeq \left[-\alpha_b+\alpha_R \hat{\textbf{z}}\times\right]\textbf{E}\int d^3r' \frac{e\mu_Bm^2k_F^2}{2\hbar^4\pi^3} e^{-k_F^2\left|\textbf{r}-\textbf{r}'\right|^2/3} \notag\\
&=\left[-\alpha_b+\alpha_R \hat{\textbf{z}}\times\right]\textbf{E}\frac{3^{\frac{3}{2}}e\mu_Bm^2}{2\hbar^4\pi^{\frac{3}{2}}k_F}. \label{Eq.diagonalME2}
\end{align}

Substituting \ref{Eq.diagonalME1} and \ref{Eq.diagonalME2} to the magnetic and electric interaction terms in \ref{Eq.Hamiltonian}, one can arrive at exchange bias Hamiltonian
\begin{align}
H_{eb}=& - \int d^3r\textbf{M}(\textbf{r})\cdot J\sum_n\textbf{M}(\textbf{r}-\textbf{r}_n)  
- \int d^3r \textbf{P}(\textbf{r})\cdot\textbf{E} \notag\\
=&\textbf{M}\cdot\textbf{B}_0,
\end{align}
where
\begin{align}
\textbf{B}_0=\left[\alpha_b-\alpha_R \hat{\textbf{z}}\times\right]\textbf{E}\frac{3^{\frac{5}{2}}e\mu_Bm^2JN}{4\hbar^4\pi^{\frac{3}{2}}k_F},
\end{align}
correspond to an exchange bias shift of the center of the magnetic hysteresis. The dependency of $\textbf{B}_0$ to $\textbf{E}$ indicates that the exchange bias is controllable by electric field. 

\begin{figure}[h]
\includegraphics[width=\columnwidth]{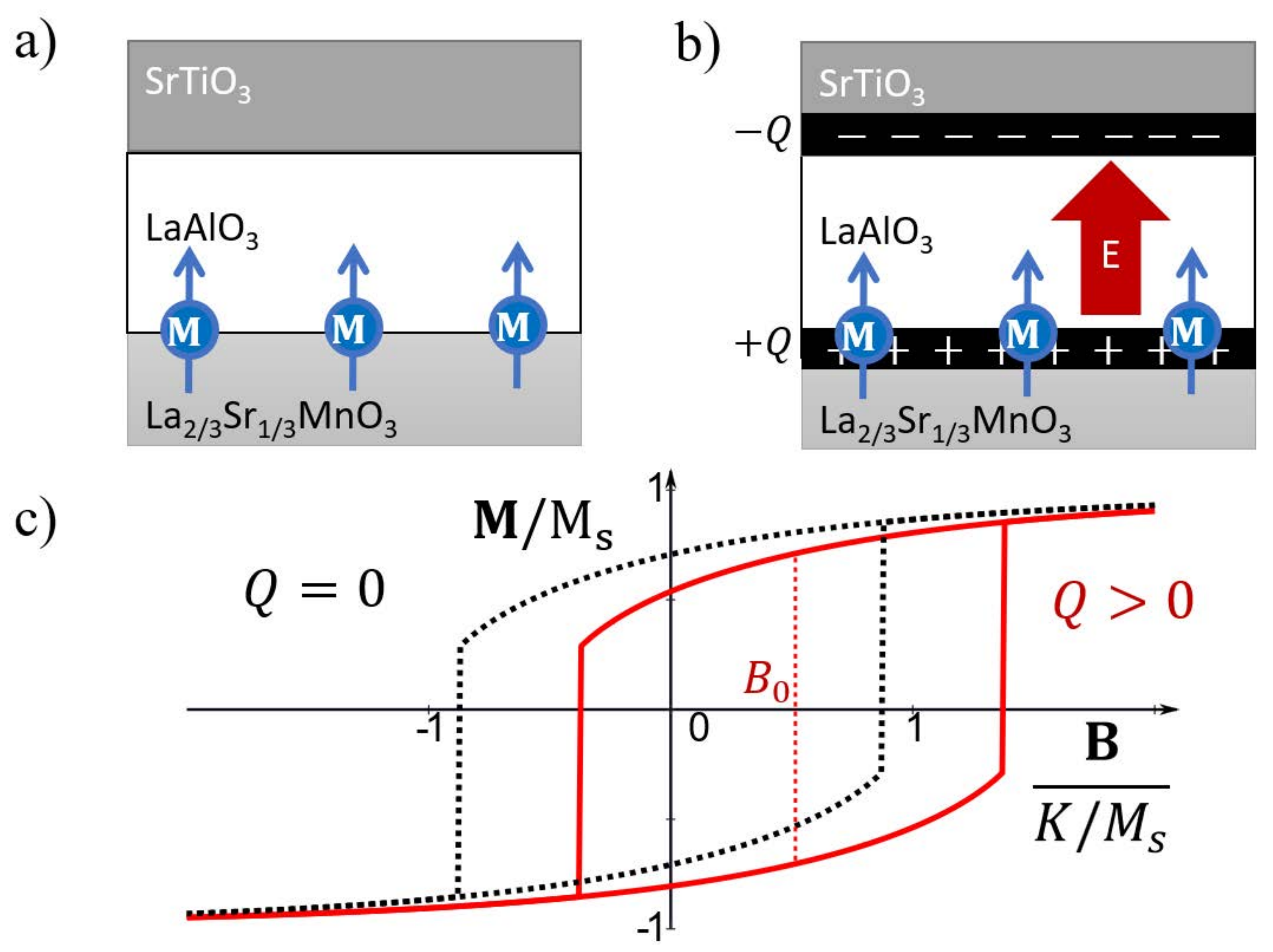}
\caption{(a) In La$_{2/3}$Sr$_{1/3}$MnO$_3|$LaAlO$_3|$SrTiO$_3$ structure, (a) Exchange bias is not observed when there is no surface charges. (b) an out-of-plane electric field generated by 2 dimensional electron gas system at the interfaces \cite{Asmara2014} induces exchange bias. (c) The exchange bias is observed as shift in the magnetic hysteresis curve. } 
\label{Fig.intE}
\end{figure}

Fig. \ref{Fig.intE} illustrates the magnetoelectric coupling of out-of-plane electric field and magnetization in a bilayer system with $\alpha_b\gg\alpha_R$. The hysteresis curve is illustrated using Stoner-Wohlfarth model with anisotropy $K$, saturation magnetization $M_S$ and angle between easy axis and field $\theta=45^\circ$. The inverse dependency of $\textbf{B}_0$ to $k_F$ is preferable for lightly-doped semiconductor, such as LaAlO$_3$ with $k_F\sim 0.3a^{-1}$, $a=3.8$\AA \ is the lattice constant \cite{anderson2023}. The out-of-plane electric field can be generated from exchange bias or from charge transfer in La$_{2/3}$Sr$_{1/3}$MnO$_3|$LaAlO$_3|$SrTiO$_3$ \cite{Asmara2014}. 
For $\alpha_bk_F=0.01$ eV, the magnitude $B_0$ can be estimated to be 
\begin{equation*}
B_0=\alpha_bE\frac{3^{\frac{5}{2}}e\mu_Bm^2JN}{4\hbar^4\pi^{\frac{3}{2}}k_F}=\alpha_b\frac{3^{\frac{3}{2}}Qe\mu_B\mu_0m^2}{2\hbar^4\pi^{\frac{3}{2}}k_Fa^2\varepsilon_r\varepsilon_0}=18 \ \mathrm{mT}
\end{equation*}
Here we used $E=Q/(\varepsilon_r\varepsilon_0 a^2)$, $\varepsilon$ is vacuum permittivity, $\varepsilon_r=25$ is the dielectric constant of LaAlO$_3$\cite{Suzuki2012}, $Q=0.5$ e \cite{Asmara2014}.
This result is in agreement with experiment by Ref.~\cite{Lu2016}, which observed that there is an exchange bias in La$_{2/3}$Sr$_{1/3}$MnO$_3|$LaAlO$_3|$SrTiO$_3$ structure when the thickness LaAlO$_3$ is more than 4 unit cell, with $B_0\sim 20$ mT. This phenomena occurs because 2 dimensional electron gas emerges at the interface of LaAlO$_3|$SrTiO$_3$ when the thickness of LaAlO$_3$ is more than 4 unit cell \cite{PhysRevLett.105.236802,Asmara2014}.

\begin{figure}[h]
\includegraphics[width=\columnwidth]{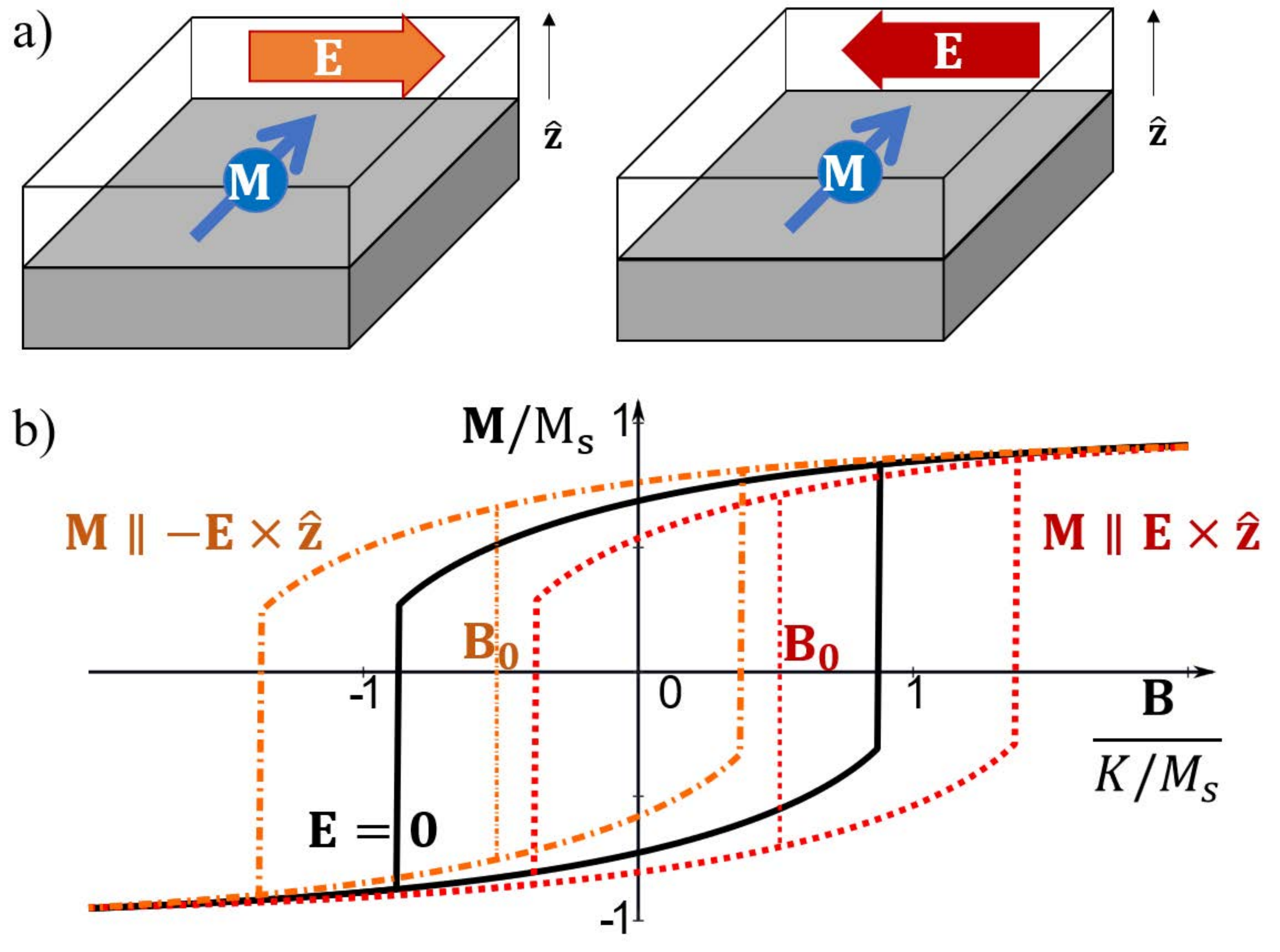}
\caption{Manipulation of exchange bias by in-plane electric field via interface spin-orbit coupling. (a) Relative directions of electric field $E$ and magnetization $\textbf{M}$ induces (b) exchange bias of the magnetic heterostructure. }
\label{Fig.extE}
\end{figure}

Fig. \ref{Fig.extE} illustrates a bilayer system with $\alpha_R\gg\alpha_b$ and an in-plane magnetization. In this case, the exchange bias of in-plane magnetization can be controlled by in-plane electric fields. The direction of electric field is perpendicular to magnetization direction. A more efficient manipulation of magnetic memory can be further developed by combining electrical control of exchange bias and spin-orbit torque \cite{ZHANG20161}, because magnetizations also manipulated by transverse electric field in spin-orbit torques devices \cite{9427163}.

\section{Summary and conclusion}
\label{SecSummary}

To summarize, we study the origin of interface magnetoelectric and its application for electrical control of exchange bias. The interface magnetoelectric arises from the spin-orbit couplings due to broken symmetry of the magnetic structure. We consider spin-orbit couplings that arise from noncentrosymmetry of the bulk structure and Rashba effect at the interface. The magnetoelectric effect associated with spin-orbit coupling due to noncentrosymmetry can describe the exchange bias observed in La$_{2/3}$Sr$_{1/3}$MnO$_3|$LaAlO$_3|$SrTiO$_3$ structure, as illustrated in Fig.~\ref{Fig.intE}. 
On the other hand, the magnetoelectric effect associated with spin-orbit coupling due to interface Rashba effect describes the coupling of in-plane magnetization with transverse electric, as illustrated in Fig.~\ref{Fig.extE}. The manipulation of exchange bias by transverse electric may be combined with spin-orbit torque to for a more efficient magnetic memory and spintronic devices.

\section*{Acknowledgement}
We thank Indonesia Toray Science Foundation for funding this research through Science \& Technology Research Grant.

$\bibliographystyle{IEEEtran}
$\bibliography{IEEEabrv,refs}


\end{document}